\documentstyle[12pt]{article}
\newcommand{\be}{\begin{equation}}
\newcommand{\ee}{\end{equation}}
\newcommand{\bea}{\begin{eqnarray}}
\newcommand{\eea}{\end{eqnarray}}
\newcommand{\bean}{\begin{eqnarray*}}
\newcommand{\eean}{\end{eqnarray*}}
\newcommand{\ba}{\begin{array}}
\newcommand{\ea}{\end{array}}

\newcommand{\slashl}[1]{\not{\!\!#1}}
\newcommand{\slashs}[1]{\not{\!#1}}
\newcommand{\norsl}{\normalsize\sl}
\newcommand{\norsc}{\normalsize\sc}
\begin{document}

\begin{titlepage}

\title{$B$ meson light-cone distribution amplitudes\\
in the heavy-quark limit}

\author{
\norsc  Hiroyuki KAWAMURA\\
\norsl  Deutsches Elektronen-Synchrotron, DESY\\
\norsl  Platanenallee 6, D 15738 Zeuthen, GERMANY\\
\norsl  and\\
\norsl  Dept. of Physics, Hiroshima University\\
\norsl  Higashi-Hiroshima 739-8526, JAPAN\\
\\
\norsc  Jiro KODAIRA\, and Cong-Feng QIAO\thanks{JSPS Research Fellow.}\\
\norsl  Dept. of Physics, Hiroshima University\\
\norsl  Higashi-Hiroshima 739-8526, JAPAN\\
\\
\norsc  Kazuhiro TANAKA\\
\norsl  Dept. of Physics, Juntendo University\\
\norsl  Inba-gun, Chiba 270-1695, JAPAN\\}

\date{}
\maketitle

\begin{abstract}
{\normalsize
\noindent
We investigate the $B$ meson light-cone distribution amplitudes
in the heavy-quark limit which are relevant for the QCD factorization
approach for the exclusive $B$ meson decays.
We derive exact relations between two- and three-particle
distribution amplitudes from the QCD equations of motion
and heavy-quark symmetry constraint.
As solution of these relations, we give representations for the
quark-antiquark distribution amplitudes in terms of independent
dynamical degrees of freedom.
In particular, we find that the Wandzura-Wilczek-type contributions
are determined uniquely in analytic form in terms of $\bar{\Lambda}$,
a fundamental mass parameter of heavy-quark effective theory,
and that both leading- and higher-twist distribution amplitudes
receive the contributions of multi-particle states with additional gluons.
}
\end{abstract}

\begin{picture}(5,2)(30,-730)
\put(0,-95){DESY 01-135}
\put(0,-110){HUPD-0108}
\put(0,-125){JUPD-0191}
\put(0,-140){September \ 2001}
\put(0,-155){Revised April \ 2002}
\put(360,-155){hep-ph/0109181 v2}
\end{picture}

\thispagestyle{empty}
\end{titlepage}
\setcounter{page}{1}
\baselineskip 18pt

Recently systematic methods based on the QCD factorization
have been developed for the exclusive $B$ meson decays
into light mesons
\cite{Beneke:2000ry,Beneke:2001wa,Pirjol:2000gn,Beneke:2001at,Bosch:2001gv}
(for other approaches see, e.g. \cite{Ball:1998kk}).
Essential ingredients in this approach are
the light-cone distribution amplitudes for the participating mesons,
which constitute nonperturbative long-distance contribution
to the factorized amplitudes \cite{Chernyak:1984ej}.
As for the light mesons ($\pi$, $K$, $\eta$, $\rho$, $\omega$, $K^{*}$,
$\phi$)
appearing in the final state,
systematic model-independent study of the light-cone distributions
exists for both leading and higher twists
\cite{Braun:1990iv,Ball:1999je,Ball:1998sk,Ball:1999ff}.
On the other hand, unfortunately,
the light-cone distribution amplitudes for
the $B$ meson are not well-known at present and
provide a major source of uncertainty in the calculations of
the decay rates.
In this Letter, we demonstrate that
heavy-quark symmetry and constraints from the equations of motion
determine a unique analytic solution for the $B$ meson light-cone
distribution amplitudes within the two-particle Fock states.
We also derive the exact integral representations
for the effects of higher Fock states with additional gluons.
A complete set of the $B$ meson distribution amplitudes
are constructed in terms of independent dynamical degrees of freedom,
which satisfies all relevant QCD constraints.

In the heavy-quark limit, the $B$ meson matrix elements obey
the heavy-quark symmetry,
and is conveniently described by the heavy-quark
effective theory (HQET) \cite{Isgur:1989vq,Neubert:1994mb}.
Following Refs.~\cite{Grozin:1997pq,Beneke:2001wa,Pirjol:2000gn},
we introduce the quark-antiquark light-cone distribution
amplitudes $\tilde{\phi}_{\pm}(t)$ of the $B$ meson
in terms of vacuum-to-meson matrix element of nonlocal
light-cone operators in the HQET:
\be
\langle 0 | \bar{q}(z) \Gamma h_{v}(0) |\bar{B}(p) \rangle
 = - \frac{i f_{B} M}{2} {\rm Tr}
 \left[ \gamma_{5}\Gamma \frac{1 + \slashs{v}}{2}
 \left\{ \tilde{\phi}_{+}(t) - \slashs{z} \frac{\tilde{\phi}_{+}(t)
 -\tilde{\phi}_{-}(t)}{2t}\right\} \right]\ .
 \label{phi}
\ee
where $z^{2}=0$, $v^{2} = 1$, $t=v\cdot z$, and $p^{\mu} = Mv^{\mu}$
is the 4-momentum of the $B$ meson with mass $M$.
$h_{v}(x)$ denotes the effective $b$-quark field,
$b(x) \approx \exp(-im_{b} v\cdot x)h_{v}(x)$,
and is subject to the on-shell
constraint, $\slashs{v} h_{v} = h_{v}$ \cite{Isgur:1989vq,Neubert:1994mb}.
$\Gamma$ is a generic Dirac matrix and,
here and in the following, the path-ordered gauge
factors are implied in between the constituent fields.
$f_{B}$ is the decay constant defined as usually as
\be
\langle 0 | \bar{q}(0) \gamma^{\mu}\gamma_{5} h_{v}(0) |\bar{B}(p) \rangle
   = i f_{B} M v^{\mu}\ ,
\label{fb}
\ee
so that $\tilde{\phi}_{\pm}(t=0) = 1$.
The behavior of the RHS of eq.(\ref{phi})
for a fast-moving meson, $t = v\cdot z \rightarrow \infty$, shows that
$\tilde{\phi}_{+}$ is the leading-twist distribution amplitude,
whereas $\tilde{\phi}_{-}$ has subleading twist~\cite{Grozin:1997pq}.

It is well-known that the equations of motion
impose a set of relations between distribution amplitudes
\cite{Braun:1990iv,Ball:1999je,Ball:1998sk,Ball:1999ff}.
To derive these relations, the most convenient method is to
start with the exact operator identities between the nonlocal operators:
\bea
 \frac{\partial}{\partial x^{\mu}}
 \bar{q}(x) \gamma^{\mu} \Gamma h_{v}(0)
  &=& \bar{q}(x) \stackrel{\leftarrow}{\slashl{D}} \Gamma h_{v}(0)
  +i \int_{0}^{1}duu \ \bar{q}(x) gG_{\mu \nu}(ux) x^{\nu}
  \gamma^{\mu}\Gamma h_{v}(0) \ ,
 \label{id1} \\
 \frac{\partial}{\partial x^{\mu}}
 \bar{q}(x) \Gamma h_{v}(0)
 &=& - \bar{q}(x) \Gamma D_{\mu} h_{v}(0)
 +i \int_{0}^{1}du (u-1)\ \bar{q}(x) gG_{\mu \nu}(ux)
 x^{\nu}\Gamma h_{v}(0) \nonumber \\
 &+& \partial_{\mu}\left\{
 \bar{q}(x) \Gamma h_{v}(0) \right\} \ ,
\label{id2}
\eea
where $x^{\mu}$ is not restricted on the light-cone.
$D_{\mu}= \partial_{\mu} - igA_{\mu}$,
$\stackrel{\leftarrow}{D}_{\mu} =
\stackrel{\leftarrow}{\partial}_{\mu} + ig A_{\mu}$
are the covariant derivatives, $G_{\mu \nu}= (i/g)[D_{\mu}, D_{\nu}]$
is the gluon field strength tensor, and
\be
  \partial_{\mu}\left\{
    \bar{q}(x) \Gamma h_{v}(0) \right\} \equiv \left.
   \frac{\partial}{\partial y^{\mu}}\
  \bar{q}(x+y) \Gamma h_{v}(y)\right|_{y \rightarrow 0}
\label{trans}
\ee
stands for the derivative over the total translation.
These identities simply describe the response of the nonlocal operators
to the change of the interquark separation and/or total translation.

By taking the vacuum-to-meson matrix element of these relations,
the operators involving the equations of motion vanish.
Going over to the light-cone limit $x_{\mu} \rightarrow z_{\mu}$,
the remaining terms can be expressed by the appropriate
light-cone distribution amplitudes $\tilde{\phi}_{+}$, $\tilde{\phi}_{-}$,
etc.
The terms given by an integral of quark-antiquark-gluon operator
are expressed by the three-particle distribution amplitudes
corresponding to the higher-Fock components of the meson wave function.
Through the Lorentz decomposition of the three-particle light-cone
matrix element, we define the four functions
$\tilde{\Psi}_V \,(t\,,\,u)$, $\tilde{\Psi}_A \,(t\,,\,u)$,
$\tilde{X}_A \,(t\,,\,u)$ and $\tilde{Y}_A \,(t\,,\,u)$
as the independent three-particle distributions:
\bea
 \lefteqn{\langle 0 | \bar{q} (z) \, g G_{\mu\nu} (uz)\, z^{\nu}
      \, \Gamma \, h_{v} (0) | \bar{B}(p) \rangle}\nonumber \\
  &=& \frac{1}{2}\, f_B M \, {\rm Tr}\, \left[ \, \gamma_5\,
      \Gamma \,
        \frac{1 + \slashs{v}}{2}\, \biggl\{ ( v_{\mu}\slashs{z}
         - t \, \gamma_{\mu} )\  \left( \tilde{\Psi}_A (t,u)
   - \tilde{\Psi}_V (t,u) \right)
      \right. \label{3elements}\\
  & & \qquad\qquad\qquad\qquad  - i \, \sigma_{\mu\nu} z^{\nu}\,
           \tilde{\Psi}_V (t,u)
       - \left. z_{\mu} \, \tilde{X}_A (t,u)\,
+ \frac{z_{\mu}}{t} \, \slashs{z} \,\tilde{Y}_A \,(t\,,\,u)
\biggr\} \, \right]\ . \nonumber
\eea
This is the most general parameterization
compatible with Lorentz invariance and the heavy-quark limit.

By using
$\bar{q}\stackrel{\leftarrow}{\slashl{D}}=0$ for the light quark
and substituting eqs.(\ref{phi}) and (\ref{3elements}),
the first identity (\ref{id1}) yields the two differential equations
connecting the two-particle distributions $\tilde{\phi}_{+}$
and $\tilde{\phi}_{-}$ with the three-particle distribution amplitudes:
\bea
 \tilde{\phi}_{-}'(t)&-& \frac{1}{t}\left(\tilde{\phi}_{+}(t)
 - \tilde{\phi}_{-}(t)\right) =
 2 t \int_0^1 du\, u\, ( \tilde{\Psi}_A (t,u) - \tilde{\Psi}_V (t,u)) \ ,
 \label{de1} \\
 \tilde{\phi}_{+}'(t)
 &-&\tilde{\phi}_{-}'(t) - \frac{1}{t}\left(\tilde{\phi_{+}}(t)
 - \tilde{\phi}_{-}(t)\right) + 4t
 \frac{\partial \tilde{\phi}_{+}(t)}{\partial z^{2}} \nonumber \\
 & & \qquad\qquad\qquad =
     2t \int_0^1 du\, u\, ( \tilde{\Psi}_A (t,u)
  + 2\, \tilde{\Psi}_V (t,u) + \tilde{X}_A (t,u)) \ ,
\label{de2}
\eea
where $\tilde{\phi}_{\pm}'(t) = d \tilde{\phi}_{\pm}(t)/dt$, and we
introduced a
shorthand notation
\be
 \frac{\partial \tilde{\phi}_{+}(t)}{\partial z^{2}} \equiv
 \left. \frac{\partial \tilde{\phi}_{+}(t, x^{2})}{\partial x^{2}}
  \right|_{x^{2} \rightarrow 0} \ .
\label{pd}
\ee
Here, via  $\tilde{\phi}_{+}(t) \rightarrow \tilde{\phi}_{+}(t, x^{2})$,
we extend the definitions in eq.(\ref{phi}) to the case $z \rightarrow x$
($x^{2} \neq 0$), since the derivative in the LHS of
eq.(\ref{id1}) has to be taken before going to the light-cone limit.

Next, we proceed to the second identity (\ref{id2}).
To use the HQET equation of motion $v \cdot D h_{v} = 0$
\cite{Isgur:1989vq,Neubert:1994mb},
we contract the both sides of eq.(\ref{id2}) with $v_{\mu}$.
After similar manipulations as above, we finally obtain
another set of two differential equations:
\bea
 \tilde{\phi}_{+}'(t)&-& \frac{1}{2t}\left(\tilde{\phi}_{+}(t)
 - \tilde{\phi}_{-}(t)\right)  + i \bar{\Lambda}\tilde{\phi}_{+}(t)
 + 2t \frac{\partial \tilde{\phi}_{+}(t)}{\partial z^{2}} \nonumber \\
  &=&  t\, \int_0^1 du\,(u-1)\,
           ( \tilde{\Psi}_A (t,u) + \tilde{X}_A (t,u) )\ , \label{de3} \\
  \tilde{\phi}_{+}'(t)
  &-&\tilde{\phi}_{-}'(t) +\left(i \bar{\Lambda} -\frac{1}{t}\right)
 \left(\tilde{\phi}_{+}(t) - \tilde{\phi}_{-}(t)\right) + 2t \left(
  \frac{\partial \tilde{\phi}_{+}(t)}{\partial z^{2}}-
  \frac{\partial \tilde{\phi}_{-}(t)}{\partial z^{2}}\right) \nonumber \\
  &=&  2t \,\int_0^1 du\,(u-1)\,
           ( \tilde{\Psi}_A (t,u) + \tilde{Y}_A (t,u) ) \ ,
\label{de4}
\eea
where
\be
  \bar{\Lambda} = M - m_{b} =
 \frac{iv\cdot \partial \langle 0| \bar{q} \Gamma h_{v} |\bar{B}(p) \rangle}
  {\langle 0| \bar{q} \Gamma h_{v} |\bar{B}(p) \rangle}
\label{lambda}
\ee
is the usual ``effective mass'' of meson states in the HQET
\cite{Falk:1992fm,Neubert:1992fk,Neubert:1994mb}.

Eqs.(\ref{de1})-(\ref{de4}) are exact in QCD
in the heavy-quark limit, and are the new results.
In the approximation that the three-particle amplitudes in the RHS
are set to zero (``Wandzura-Wilczek approximation''),
the system of differential equations (\ref{de1}), (\ref{de2}) is
equivalent to that found in Ref.~\cite{Beneke:2001wa}.
By combining eqs.(\ref{de2}) and (\ref{de3}), we can eliminate the term
$\partial \tilde{\phi}_{+} (t)/\partial z^2$ as
\be
 \tilde{\phi}_{+}' (t) + \tilde{\phi}_{-}' (t)
  + 2 i \bar{\Lambda} \tilde{\phi}_+ (t)
  = - 2 t \int_0^1 du\, \left(
    \tilde{\Psi}_A (t,u) + \tilde{X}_A (t,u) + 2 u\,
   \tilde{\Psi}_V (t,u) \right)   \ .\label{fresult5}
\ee
Now, from a system of equations (\ref{de1}) and (\ref{fresult5}),
we can obtain the solution for $\tilde{\phi}_{+}$ and $\tilde{\phi}_{-}$.
For this purpose, it is convenient to work with the momentum-space
distribution amplitudes $\phi_{\pm}(\omega)$ defined by
\be
 \tilde{\phi}_{\pm}(t) = \int d\omega \ e^{-i \omega t}
  \phi_{\pm}(\omega) \ . \label{mom}
\ee
Here $\omega v^{+}$ has the meaning of the light-cone projection
$k^{+}$ of the light-antiquark momentum in the $B$ meson.
The singularities in the complex $t$-plane
are such that $\phi_{\pm}(\omega)$ vanish for $\omega < 0$.
Similarly for the three-particle distribution amplitudes, we define
\be
 \tilde{F}(t, u) = \int d\omega d \xi \
  e^{-i(\omega  + \xi u)t} F(\omega, \xi) \ ,
   \qquad (F=\{ \Psi_{V}, \Psi_{A}, X_{A}, Y_{A}  \})\ . \label{ff3}
\ee
Here $\omega v^{+}$ and $\xi v^{+}$ denote
the light-cone projection
of the momentum carried by the light antiquark and the gluon,
respectively, and $F(\omega, \xi)$ vanishes unless
$\omega \ge 0$ and $\xi \ge 0$.
By going over to the momentum space, we obtain from
eqs.(\ref{de1}) and (\ref{fresult5}),
\bea
 \omega \frac{d \phi_{-}(\omega)}{d \omega}
  &+& \phi_{+}(\omega) = I(\omega)\ ,
  \label{mde1} \\
  \left(\omega - 2 \bar{\Lambda}\right)\phi_{+}(\omega)
 &+& \omega \phi_{-}(\omega) = J(\omega) \ , \label{mde2}
\eea
where $I(\omega)$ and $J(\omega)$ denote the
``source'' terms due to three-particle amplitudes as
\bea
 I(\omega) &=& 2\frac{d}{d\omega}
   \int_{0}^{\omega}d\rho \int_{\omega - \rho}^{\infty}\frac{d\xi}{\xi}
  \frac{\partial}{\partial \xi}\left[ \Psi_{A}(\rho, \xi)
   - \Psi_{V}(\rho, \xi)\right] \ , \label{si} \\
 J(\omega) &=& -2\frac{d}{d\omega}
  \int_{0}^{\omega}d\rho \int_{\omega - \rho}^{\infty}\frac{d\xi}{\xi}
  \left[ \Psi_{A}(\rho, \xi) + X_{A}(\rho, \xi)\right]
  \nonumber \\
  && -4 \int_{0}^{\omega}d\rho \int_{\omega - \rho}^{\infty}\frac{d\xi}{\xi}
  \frac{\partial \Psi_{V}(\rho, \xi)}{\partial \xi} \ . \label{sj}
\eea
We solve eqs.(\ref{mde1}) and (\ref{mde2}) with
boundary conditions $\phi_{\pm}(\omega) = 0$
for $\omega < 0$ or $\omega \rightarrow \infty$.
Obviously, the solution can be decomposed into two pieces as
\be
  \phi_{\pm}(\omega) = \phi_{\pm}^{(WW)}(\omega)
  + \phi_{\pm}^{(g)}(\omega) \ ,
\label{decomp}
\ee
where $\phi_{\pm}^{(WW)}(\omega)$ are the solution of
eqs.(\ref{mde1}) and (\ref{mde2}) with the source terms set to zero,
$I(\omega)=J(\omega)=0$, i.e., in the Wandzura-Wilczek approximation.
$\phi_{\pm}^{(g)}(\omega)$ denote the pieces induced by the source terms.

First, let us discuss $\phi_{\pm}^{(WW)}$.
Eq.(\ref{mde1}) alone, with $I(\omega)=0$, is
equivalent to a usual Wandzura-Wilczek type relation derived
in Ref.~\cite{Beneke:2001wa}:
\be
 \phi_{-}^{(WW)}(\omega) = \int_{\omega}^{\infty}
  d\rho \ \frac{\phi_{+}^{(WW)}(\rho)}{\rho} \ .
\label{ww}
\ee
Now combining eqs.(\ref{mde1}) and (\ref{mde2}), we are able to obtain
the analytic solution explicitly as
\bea
 \phi_{+}^{(WW)}(\omega) &=& \frac{\omega}{2 \bar{\Lambda}^{2}}
 \theta(2 \bar{\Lambda} - \omega) \ , \label{solp} \\
 \phi_{-}^{(WW)}(\omega) &=&
  \frac{2 \bar{\Lambda} - \omega}{2 \bar{\Lambda}^{2}}
  \theta(2 \bar{\Lambda} - \omega) \ , \label{solm}
\eea
for $\omega \ge 0$.
These are normalized as
$\int_{0}^{\infty}d\omega \phi_{\pm}^{(WW)}(\omega) = 1$.
They vanish for $\omega > 2\bar{\Lambda}$, and we note that
$2\bar{\Lambda}$ is actually the kinematical upper bound of $\omega$
allowed for the two-particle Fock states of the $B$ meson
in the heavy quark limit.

The solution for $\phi_{\pm}^{(g)}$ can be obtained
straightforwardly, and reads:
\bea
 \phi_{+}^{(g)}(\omega) &=& \frac{\omega}{2\bar{\Lambda}}\Phi(\omega)\ ,
  \label{solpg} \\
  \phi_{-}^{(g)}(\omega) &=&
  \frac{2\bar{\Lambda}-\omega}{2\bar{\Lambda}}\Phi(\omega)
 + \frac{J(\omega)}{\omega}\ ,\label{solmg}
\eea
where $\omega \ge 0$ and
\bea
 \Phi(\omega) &=& \theta(2\bar{\Lambda}-\omega)
 \left\{\int_{0}^{\omega}d\rho \frac{K(\rho)}{2\bar{\Lambda} - \rho}
 -\frac{J(0)}{2\bar{\Lambda}}\right\}
 - \theta(\omega - 2\bar{\Lambda}) \int_{\omega}^{\infty}
  d\rho \frac{K(\rho)}{2\bar{\Lambda} - \rho} \nonumber \\
   &-& \int_{\omega}^{\infty}d\rho
  \left( \frac{K(\rho)}{\rho} + \frac{J(\rho)}{\rho^{2}} \right) \ ,
\label{Phi}
\eea
with
\be
  K(\rho) = I(\rho) + \left( \frac{1}{2\bar{\Lambda}} -
    \frac{d}{d\rho}\right)J(\rho)\ .
\label{K}
\ee
These functions obey
$\int_{0}^{\infty}d\omega \phi_{\pm}^{(g)}(\omega) = 0$, so that
the total amplitudes are normalized as
$\int_{0}^{\infty}d\omega \phi_{\pm}(\omega)= \tilde{\phi}_{\pm}(0)= 1$.
The solution (\ref{decomp}) with eqs.(\ref{solp})-(\ref{K})
for the distribution amplitudes is exact and
presents the principal result of our paper.
Eq.(\ref{decomp}) might exhibit discontinuity at $\omega = 2\bar{\Lambda}$,
but this does not constitute a problem because physical observables are
given by convolution integrals of distribution amplitudes
with smooth coefficient functions, i.e., the sharp behavior is
averaged over.
Namely, the distribution amplitudes should generally be understood
as distributions (in the mathematical sense).

In terms of the Mellin moments ($n= 0, 1, 2, \cdots$),
\bea
 \langle \omega^{n} \rangle_{\pm} = \int_{0}^{\infty} d\omega \ \omega^{n}
  \phi_{\pm}(\omega)
  &=&\int_{0}^{\infty} d\omega \ \omega^{n}
  \phi_{\pm}^{(WW)}(\omega) + \int_{0}^{\infty} d\omega \ \omega^{n}
  \phi_{\pm}^{(g)}(\omega)  \nonumber \\
  &\equiv& \langle \omega^{n} \rangle_{\pm}^{(WW)}
   +\langle \omega^{n} \rangle_{\pm}^{(g)}\ , \label{mel}
\eea
our solution reads:
\be
 \langle \omega^{n} \rangle_{+}^{(WW)}= \frac{2}{n+2}(2\bar{\Lambda})^{n} \
,
   \qquad \langle \omega^{n} \rangle_{-}^{(WW)}=
  \frac{2}{(n+1)(n+2)}(2\bar{\Lambda})^{n} \ ,
\label{melsol}
\ee
\bea
 \langle \omega^{n} \rangle_{+}^{(g)}
  &=& \frac{2}{n+2}\sum_{i=1}^{n-1}(2\bar{\Lambda})^{i-1}
  \sum_{j=1}^{n-i} {n-i \choose j}
 \left( \left\{(n+1-i)\frac{2j+1}{j+1} + 1\right\}
  \left[\Psi_{A}\right]^{n-i}_{j} \right. \nonumber\\
   &+&\left. (n + 2 - i) \left[X_{A}\right]^{n-i}_{j}
   + (n + 3 - i)\frac{j}{j+1} \left[\Psi_{V}\right]^{n-i}_{j} \right)\ ,
  \label{melpg} \\
 \langle \omega^{n} \rangle_{-}^{(g)}
  &=&\frac{1}{n+1}\langle \omega^{n} \rangle_{+}^{(g)}
  - \frac{2n}{n+1}\sum_{j=1}^{n-1}{n-1 \choose j}\frac{j}{j+1}
  \left( \left[\Psi_{A}\right]^{n-1}_{j} -\left[\Psi_{V}\right]^{n-1}_{j}
 \right) \ . \label{melmg}
\eea
where ${i \choose j} = i!/[j!(i-j)!]$,
and we introduced the double moments of the three-particle distributions as
\be
 \left[F\right]^{i}_{j} =\int_{0}^{\infty}d\omega \int_{0}^{\infty}d\xi \
 \omega^{i-j} \xi^{j-1}
  F(\omega, \xi)\ ,\qquad (F=\{ \Psi_{V}, \Psi_{A}, X_{A}\})\ .
\label{dm}
\ee
For a few low moments, eqs.(\ref{mel})-(\ref{dm}) give
\bea
 \langle \omega \rangle_{+} &=& \frac{4}{3}\bar{\Lambda}\ ,
 \qquad \qquad \qquad \qquad \langle \omega \rangle_{-}=
  \frac{2}{3}\bar{\Lambda} \ ,
 \label{mome1} \\
\langle \omega^{2} \rangle_{+} &=& 2
\bar{\Lambda}^{2}+\frac{2}{3}\lambda_{E}^{2}
 +\frac{1}{3}\lambda_{H}^{2} \ ,\quad\
 \langle \omega^{2} \rangle_{-} = \frac{2}{3} \bar{\Lambda}^{2}
 +\frac{1}{3}\lambda_{H}^{2} \ .
\label{mome2}
\eea
Here we have used
\be
 \left[\Psi_{A}\right]_{1}^{1} = \frac{1}{3}\lambda_{E}^{2}\ ,
 \quad \left[\Psi_{V}\right]_{1}^{1} = \frac{1}{3}\lambda_{H}^{2}\ ,
 \quad \left[X_{A}\right]^{1}_{1} = 0\ ,
\label{mome11}
\ee
where $\lambda_{E}$ and $\lambda_{H}$ parameterize the two independent
reduced
matrix elements of local quark-antiquark-gluon operators of dimension 5,
and are related to the chromoelectric and chromomagnetic fields
in the $B$ meson rest frame~\cite{Grozin:1997pq},
\bea
 \langle 0 |\bar{q} g \mbox{\boldmath $E$}\cdot\mbox{\boldmath $\alpha$}
 \gamma_{5}h_{v} |\bar{B}(\mbox{\boldmath $p$}=0)\rangle
 &=& f_{B}M \lambda_{E}^{2} \ ,
 \label{lambdae}\\
 \langle 0 |\bar{q} g \mbox{\boldmath $H$}\cdot\mbox{\boldmath $\sigma$}
 \gamma_{5}h_{v} |\bar{B}(\mbox{\boldmath $p$}=0)\rangle
 &=& if_{B}M \lambda_{H}^{2} \ ,
 \label{lambdah}
\eea
with $E^{i}=G^{0i}$, $H^{i}=-\frac{1}{2}\epsilon^{ijk}G^{jk}$, and
$\mbox{\boldmath $\alpha$}= \gamma^{0}\mbox{\boldmath $\gamma$}$.
The results (\ref{mome1}) and (\ref{mome2}) exactly coincide with the
relations
obtained by Grozin and Neubert \cite{Grozin:1997pq},
who have derived their relations by analyzing matrix elements of some
{\it local} operators.
Our results (\ref{mel})-(\ref{dm}) from nonlocal operators
give generalization of theirs to $n \ge 3$.

An interesting feature revealed by our results is that
the leading-twist distribution amplitude $\phi_{+}$ as well as
the higher-twist $\phi_{-}$
contains the three-particle contributions.
This is in contrast with the case of the light mesons,
where the leading-twist amplitudes correspond to
the ``valence'' Fock component of the wave function,
while the higher-twist amplitudes involve contributions of
multi-particle states.

In the higher-twist distribution amplitudes of light mesons
\cite{Braun:1990iv,Ball:1999je,Ball:1998sk,Ball:1999ff},
the contributions of multi-particle states
with additional gluons have been generally important and broadened the
distributions, but they have typically
produced corrections less than $\sim 20$\% to the main term
given by the Wandzura-Wilczek contributions.
We note that, in the present case, there exists
a rough estimate
$\lambda_{E}^{2}/\bar{\Lambda}^{2} = 0.36 \pm 0.20$,
$\lambda_{H}^{2}/\bar{\Lambda}^{2} = 0.60 \pm 0.23$
by QCD sum rules \cite{Grozin:1997pq} (see eq.(\ref{mome2})),
but any estimate of the higher moments is not known.
It is obvious that further investigations are required
to clarify the effects of multi-particle states.
In the applications to the physical amplitudes,
the evolution effects including the three-body operators also enter the
game.
All these further developments for going beyond the Wandzura-Wilczek
approximation
can be exploited systematically starting from the exact results in this
paper,
as it has been done for light mesons.

Inspired by the QCD sum rule estimates, Grozin and Neubert
\cite{Grozin:1997pq}
have proposed model distribution amplitudes
$\phi^{GN}_{+}(\omega) = \left(\frac{\omega}{\omega_{0}^{2}}\right)
\exp \left(- \frac{\omega}{\omega_{0}}\right)$,
$\phi^{GN}_{-}(\omega) = \left(\frac{1}{\omega_{0}}\right)
\exp \left(-\frac{\omega}{\omega_{0}}\right)$,
where $\omega_{0} = 2\bar{\Lambda}/3$.
The shape of their model distributions is rather different from that
of the Wandzura-Wilczek
contributions (\ref{solp}) and (\ref{solm}), except
the behavior $\phi_{+}^{GN}(\omega) \sim \omega$,
$\phi_{-}^{GN}(\omega) \sim {\rm const}$,
as $\omega \rightarrow 0$.
Actually, such behavior when the light-antiquark becomes ``soft''
is suggested by the corresponding behavior
of the light-mesons \cite{Braun:1990iv,Ball:1999je,Ball:1998sk,Ball:1999ff}.
But we note that the gluon correction (\ref{solpg}) to $\phi_{+}(\omega)$
would modify such behavior if
$J(0) = -2 \int_{0}^{\infty}(d\xi/\xi)\left(\Psi_{A}(0, \xi)
 + X_{A}(0, \xi)\right) \neq 0$.
\footnote{$\phi_{-}^{(g)}(0)$ of eq.(\ref{solmg})
is finite even if $J(0) \neq 0$.}

To summarize, the solution in this paper
provides the powerful framework for building up the $B$ meson
light-cone distribution amplitudes and their phenomenological applications.
Our results represent
the quark-antiquark distribution amplitudes
in terms of independent dynamical degrees of freedom,
and satisfy the constraints from the equations of motion exactly.
The other essential constraints are imposed by
heavy-quark symmetry, which lead to the reduced set of the
distribution amplitudes and thus allow us to determine the Wandzura-Wilczek
contributions explicitly in analytic form.
The Wandzura-Wilczek contributions give the
effects corresponding to the valence distributions, while the corrections
due to
the additional gluons are also obtained as exact integral representations
involving the three-particle distribution amplitudes.
A detailed study of the gluon corrections requires systematic treatment of
the
three-particle distributions, and will be presented elsewhere.

We have treated the light-cone
distribution amplitudes for the pseudoscalar
$B$ meson. In the heavy-quark limit,
the distribution amplitudes of the vector meson $B^{*}$ is related with
those of the $B$ meson thanks to heavy-quark spin symmetry
\cite{Grozin:1997pq},
so that the solution in this paper also determines
a complete set of the $B^{*}$ meson distribution amplitudes.

Among the four differential equations (\ref{de1})-(\ref{de4})
derived in this paper,
we have utilized only a system of two equations to obtain
the solution (\ref{decomp}).
Now, with our explicit solution,
the other two equations can be used to determine
$\partial \tilde{\phi}_{\pm} (t)/\partial z^2$ (see eq.(\ref{pd})),
i.e., the transverse momentum dependence of the distributions,
which are necessary for computing the power corrections
to the exclusive amplitudes.

\vspace{0.5cm}
The authors would like to thank T. Onogi for fruitful discussions.
The work of J.K. was supported in part by the Monbu-kagaku-sho Grant-in-Aid
for Scientific Research No.C-13640289.
The work of C-F.Q. was supported by the Grant-in-Aid of JSPS committee.


\newpage
\begin{center}
{\large\bf Erratum to: ``$B$ meson light-cone distribution amplitudes\\
in the heavy-quark limit''}\\
{\large\bf [Phys. Lett. B 523 (2001) 111]}
\end{center}
\begin{center}
{\normalsize
Hiroyuki Kawamura\ $^{\rm a}$,
Jiro Kodaira\ $^{\rm b}$, Cong-Feng Qiao\ $^{\rm b}$, 
Kazuhiro Tanaka\ $^{\rm c}$}
\end{center}

\begin{center}
{\footnotesize
$^{\rm a}${\it DESY-Zeuthen, Germany}\\
$^{\rm b}${\it Department of Physics, Hiroshima University, Japan}\\
$^{\rm c}${\it Department of Physics, Juntendo University, Japan}
}
\end{center}


%
%
\baselineskip 18pt

\bigskip

There was an error in Eq. (6) of our Letter.
The correct equation is 
\setcounter{equation}{5}
\bea
 \lefteqn{\langle 0 | \bar{q} (z) \, g G_{\mu\nu} (uz)\, z^{\nu}
      \, \Gamma \, h_{v} (0) | \bar{B}(p) \rangle}\nonumber \\
  &=& \frac{1}{2}\, f_B M \, {\rm Tr}\, \left[ \, \gamma_5\,
      \Gamma \,
        \frac{1 + \slashs{v}}{2}\, \biggl\{ ( v_{\mu}\slashs{z}
         - t \, \gamma_{\mu} )\  \left( \tilde{\Psi}_A (t,u)
   - \tilde{\Psi}_V (t,u) \right)
      \right. \label{3elements0}\\
  & & \qquad\qquad\qquad\qquad  - i \, \sigma_{\mu\nu} z^{\nu}\,
           \tilde{\Psi}_V (t,u)
       - \left. z_{\mu} \, \tilde{X}_A (t,u)\,
+ \frac{z_{\mu}}{t} \, \slashs{z} \,\tilde{Y}_A \,(t\,,\,u)
\biggr\} \, \right]\ , \nonumber
\eea
which involves the four functions
$\tilde{\Psi}_V \,(t\,,\,u)$, $\tilde{\Psi}_A \,(t\,,\,u)$,
$\tilde{X}_A \,(t\,,\,u)$ and $\tilde{Y}_A \,(t\,,\,u)$
as the independent three-particle distribution amplitudes.
Due to this change, Eq. (11) must be also replaced by
\setcounter{equation}{10}
\bea
  \tilde{\phi}_{+}'(t)
  &-&\tilde{\phi}_{-}'(t) +\left(i \bar{\Lambda} -\frac{1}{t}\right)
 \left(\tilde{\phi}_{+}(t) - \tilde{\phi}_{-}(t)\right) + 2t \left(
  \frac{\partial \tilde{\phi}_{+}(t)}{\partial z^{2}}-
  \frac{\partial \tilde{\phi}_{-}(t)}{\partial z^{2}}\right) \nonumber \\
  &=&  2t \,\int_0^1 du\,(u-1)\,
           ( \tilde{\Psi}_A (t,u) + \tilde{Y}_A (t,u) ) \ .
\label{de40}
\eea
All other equations and the conclusions presented in the Letter remain unchanged.
\end{document}